\documentclass[twocolumn,10pt,nofootinbib,preprintnumbers]{revtex4-1}

\usepackage[linewidth=1pt]{mdframed}
\usepackage[margin=0.8in]{geometry}
\usepackage[english]{babel}
\usepackage[utf8]{inputenc}
\usepackage{amsmath}
\usepackage{graphicx}
\usepackage{physics,amssymb}
\usepackage{hyperref}
\usepackage{array}
\hypersetup{
}
\usepackage{xparse}

\usepackage[normalem]{ulem}

\newcommand\dsum{\sum\limits}

\definecolor{orange}{RGB}{222, 102, 13}

\definecolor{crimson}{RGB}{162, 0, 37}
\newcommand{\change}[1]{\textcolor{crimson}{#1}}

\newcommand{\beq}{\begin{equation}}% can be used as {equation} or {eqnarray}
\newcommand{\eeq}{\end{equation}}
\newcommand{\bea}{\begin{eqnarray}} % can be used as {equation} or {eqnarray}
\newcommand{\eea}{\end{eqnarray}}
\newcommand{\nn}{\nonumber}
\def\beqa{\begin{eqnarray}}
\def\eeqa{\end{eqnarray}}

\usepackage{makecell}
\newcolumntype{?}{!{\vrule width 0.1pt}}

\begin{document}

\title{Roadmap to Thermal Dark Matter Beyond the WIMP Unitarity Bound} 
\author{Ronny Frumkin}\email{ronny.frumkin@mail.huji.ac.il}
\author{Eric Kuflik}\email{eric.kuflik@mail.huji.ac.il}
\author{Itay Lavie}\email{itay.lavie@mail.huji.ac.il}
\author{Tal Silverwater}\email{tal.silverwater@mail.huji.ac.il}
\affiliation{Racah Institute of Physics, Hebrew University of Jerusalem, Jerusalem 91904, Israel}
\date{\today}

\begin{abstract}
 We study the general properties of the freezeout of a thermal relic. We give analytic estimates of the relic abundance for an arbitrary freezeout process, showing when instantaneous freezeout is appropriate and how it can be corrected when freezeout is slow. This is used to generalize the relationship between the dark matter mass and coupling that matches the observed abundance. The result encompasses well-studied particular examples, such as WIMPs, SIMPs, coannihilation, coscattering, inverse decays, and forbidden channels, and generalizes beyond them.  In turn, this gives an approximate perturbative unitarity bound on the dark matter mass for an arbitrary thermal freezeout process. We show that going beyond the maximal masses allowed for freezeout via dark matter self-annihilations (WIMP-like, $m_{\rm DM}\gg\mathcal{O}(100~\rm TeV)$) predicts that there are nearly degenerate states with the dark matter  and that the dark matter is generically metastable. 
\end{abstract}

\maketitle

\flushbottom

\section{Introduction}
\label{sec:intro}
Overwhelming evidence, accumulating from the beginning of the last century, suggest the existence of dark matter (DM), yet we still don't know much about it. All we know is inferred from its gravitational interactions, while its particle nature remains a mystery: we still don't know its mass, forces it interacts with, and most importantly - how we may discover it.

We expect that the current DM abundance can be related to its microscopic properties such as mass and interaction cross-section, analogous  for instance, to how the abundances of the light nuclear elements are determined~\cite{Alpher:1948ve,Cyburt:2015mya}. Furthermore, the possibility that the DM is a thermal relic is particularly exciting, namely that the DM interactions brought it to equilibrium with the Big Bang bath, and its interactions determine its abundance~\cite{Lee:1977ua,Scherrer:1985zt}. Being a thermal relic, the abundance today will be insensitive to any initial conditions before the equilibrium state.   Thermal relics are also very predictive, since the relic abundance is determined by size of the dominant interaction during freezeout.

The dominant paradigm of DM has been Weakly-Interacting-Massive-Particles (WIMPs), whose departure from equilibrium is determined by DM-DM annihilations into light Standard Model (SM) bath particles.\footnote{Recent years have brought forth a number of new thermal processes  controlling the DM abundance beyond ordinary DM-DM annihilations, including~\cite{Hambye:2008bq,DEramo:2010keq,Hochberg:2014dra,Kuflik:2015isi,Dey:2016qgf,Cline:2017tka,
DAgnolo:2017dbv,Garny:2017rxs,Smirnov:2020zwf,Kramer:2020sbb,Fitzpatrick:2020vba,Frumkin:2021zng}.}
This paradigm has led to the belief that a thermal elementary DM candidate, within standard cosmological history, has an upper bound on its mass of $\sim 100~{\rm TeV}$~\cite{Griest:1989wd}. However recent works~\cite{Kim:2019udq,Kramer:2020sbb} have shown number of exceptions (for heavy dark matter in other scenarios see the review in \cite{Carney:2022gse} and references within).  In this work we put these examples in a general framework and derive bounds for a family of such mechanism. 

In section~\ref{sec:freezeout}, we present a general framework to study thermal freezeout processes,   focusing on   mechanisms which provide DM masses higher then the WIMP unitarity bound (superheavy). We emphasize the characterizing properties like slow freezeout and solve for the mass - coupling relation showing excellent agreement with numeric results.

In section~\ref{sec:unitarity}, we use the above results and derive an approximate pertubative unitarity bound for a general process. We show that a  consequence of a DM candidate $\chi$  beyond the maximal mass allowed by the WIMP unitarity bound %for freezeout via dark matter self-annihilations 
% We show that going beyond the maximal masses allowed by unitarity for freezeout via dark matter self-annihilations imposes %the constraint
% $
% m_\chi \le \sum_i m_i+\sum_f m_f < 3 m_\chi,
% $
% where $\sum_i m_i+\sum_f m_f $  is the sum of masses of all the initial and final state particles, excluding one initial dark matter particle. 
% %\change{where $\sum_j m_j$  is the sum of masses of all particles in the process but one DM particle.}
%  This is based on  requiring that the DM is stable against immediate decay to the other particles in the interaction, and requiring the DM is indeed superheavy.%This condition can be strengthened for heavier mass bounds up to the natural cut-off of $m_{\rm pl}$. \ek{I dont like this last sentence}
%An immediate result from \eqref{massboundintro} is
is that there is another particle close to the mass of $\chi$, which we will call $\chi'$. If the mass of $\chi'$ is above the WIMP unitarity bound, then there must be a nearly degenerate particle to $\chi'$ as well, resulting in a chain of superheavy particles. The chain can end only either when we reach a particle below the WIMP unitarity bound or with an unstable particle.
Such cases are exemplified in scenarios  within a standard cosmological history~\cite{Kim:2019udq,Kramer:2020sbb} and within an early matter dominated scenario~\cite{Berlin:2017ife}. Finally, we elaborate on this concept and analyze two examples of chains in section~\ref{sec:chain}.

\section{Thermal Freezeout }
\label{sec:freezeout}

\begin{table*}
\begin{minipage}{\textwidth}	
\renewcommand{\arraystretch}{1.35}
\centering
\begin{tabular}{l||lllll}
                & Process                                             & Masses                                  & $b$         & ~~~~~$\beta-1$ & Unitarity bound (GeV) \\ \hline\hline
WIMP            & $ \chi \chi \to \phi \phi$                      & $m_\phi \ll m_\chi$                     & 1           &      $\frac{1}{x_d} \log(1 + x_d)$   & $\mathcal{O}(10^5)$     \\\Xhline{0.25\arrayrulewidth}
SIMP            & $\chi \chi \chi \to \chi \chi$~~                      &                                         & 2           & $\frac{1}{2 x_d} \log(1 + x_d
/2)$    & $\mathcal{O}(1)$   \\\Xhline{0.25\arrayrulewidth}
Forbidden       & $\chi \chi \to \psi \psi$                           & $m_\psi =m_\chi(1+\Delta)$              & $1+2\Delta ~~$ & $\frac{1}{x_d} \log(1 + x_d)$ & $\mathcal{O}(10^{5-14 \frac{\Delta}{1+\Delta}})$        \\\Xhline{0.25\arrayrulewidth}
Coannihilations & \!\!$\begin{array}{l}\psi \psi \to \phi \phi \\ \chi \chi \leftrightarrow \psi \psi \end{array}$~~ & $m_\psi =m_\chi(1+\Delta)$              & $1+\Delta$~~  &   $\frac{1}{x_d} \log(1 + x_d)$  & $\mathcal{O}(10^{5-7 \frac{\Delta}{1+\Delta/2}})$     \\\Xhline{0.25\arrayrulewidth}
Coscattering    & $\chi \phi\to \psi \phi$                        & $m_\phi \ll m_\psi =m_\chi(1+\Delta)$ & $\Delta$    &    $\begin{cases} 
  \frac{1}{\Delta x_d} & {\rm for~} c=1	{\rm ~or~} \Delta\gg 0 \\ 
 \frac{1}{c-2} & {\rm for~} c>2 {\rm ~and~}  \Delta=0 \\
 \end{cases}$  ~~~~& $\mathcal{O}(10^{9\frac{2-\Delta}{1+\Delta}})$   \\\Xhline{0.25\arrayrulewidth}
Zombie          & $\chi \psi \to \psi \psi$                           & $m_\psi =m_\chi(1-\Delta)$              & $1-\Delta$  &         $\frac{1}{(1-\Delta) x_d }$ &$\mathcal{O}(10^{5+7 \frac{\Delta}{1-\Delta/2}})$  \\\Xhline{0.25\arrayrulewidth}
Inverse decays  & $\chi \phi \to \psi$                              & $m_\phi \ll m_\psi =m_\chi(1+\Delta)$ ~~ & $\Delta$    &        $\frac{1}{\Delta x_d }\left(1+\frac{1}{\Delta x_d }\right)  $ & $\mathcal{O}(10^{9\frac{2-\Delta}{1+\Delta}})$
\end{tabular}
\caption{Parameters and the maximum mass allowed by perturbative unitarity for different freezeout processes. Here $\chi$ is the DM, $\psi$ is particle close to mass with $\chi$, and $\phi$ is a particle much lighter than both $\chi$ and $\psi$. The definition of $\Delta$ is implicitly defined in the Masses column for each process. Both $\psi$ and $\phi$ are assumed to be in thermal and chemical equilibrium throughout the freezeout process. In the $2 \to 2$ freezeout cases, we assumed s-wave annihilations for calculating $\beta$.   For coannihilations, the process $\psi \psi \to \phi \phi $ controls the abundance, where $\chi \chi \leftrightarrow \psi \psi $ is assumed to be in  equilibrium throughout freezeout. }
\label{tablesigma}
\end{minipage}
\end{table*}

In this section we consider a DM candidate in thermal equilibrium with the bath and in chemical equilibrium via the  $n\to m$ process of the form 
\beq \label{process}
\underbrace{\chi\,\chi\,\cdots \chi}_{k}\, i_1 \, i_2\cdots i_{n-k} \to \underbrace{\chi\,\chi\,\cdots \chi }_{\ell}\,
f_1 \, f_2\cdots \,f_{m-\ell}. 
\eeq
The general form of the Boltzmann equation for the $\chi$ number density is
\beq \label{boltzn}
\dot{n}_{\chi}+3 H n_{\chi}=-\langle\sigma v\rangle (n_{i_1}^{\rm eq}\cdots n_{i_{n-k}}^{\rm eq})
\left(n_{\chi}^{k}-n_\chi^{\ell} \left(n_{\chi}^{eq}\right)^{k-\ell}\right), 
\eeq
neglecting Pauli-blocking and stimulated emission. 
Here $\langle\sigma v\rangle$ is the thermally averaged cross-section of the $n \to m$ process in Eq.~\eqref{process}. On dimensional grounds, this thermally averaged cross section will have the form
\beq
\langle\sigma v\rangle \equiv  \frac{\alpha_{\rm eff} ^{m+n-2}}{m_\chi^{3 n-4}} \times \frac{e^{-b'x}}{x^{c'}} \,. \label{xs-param}
\eeq
for some  coupling $\alpha_{\rm eff}$, and parameters $b'$ and $c'$. Here we define the dimensionless time parameter, $x={m_\chi}/{T}$. The power of $\alpha_{\rm eff} $ is chosen to be the form expected if the amplitude is determined solely by cubic interactions, such as gauge or Yukawa interactions, but this need not be the case.  Usually,  the temperature dependence can be well approximated at freezeout  by a power law  or exponential dependence on the temperature, which occurs for instance for forbidden channels.  

\begin{figure*}[t!]
    \centering
\includegraphics[width=.967\columnwidth]{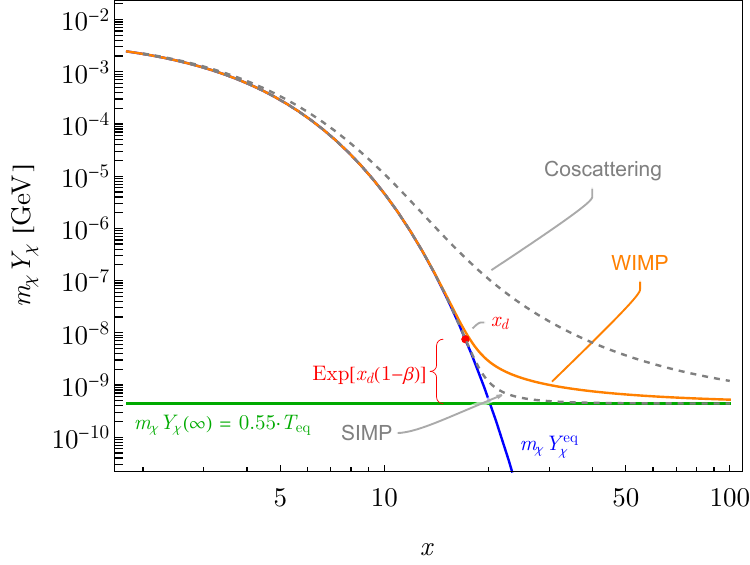}
  \vspace{-0.7cm}\hfill
\includegraphics[width=1.\columnwidth]{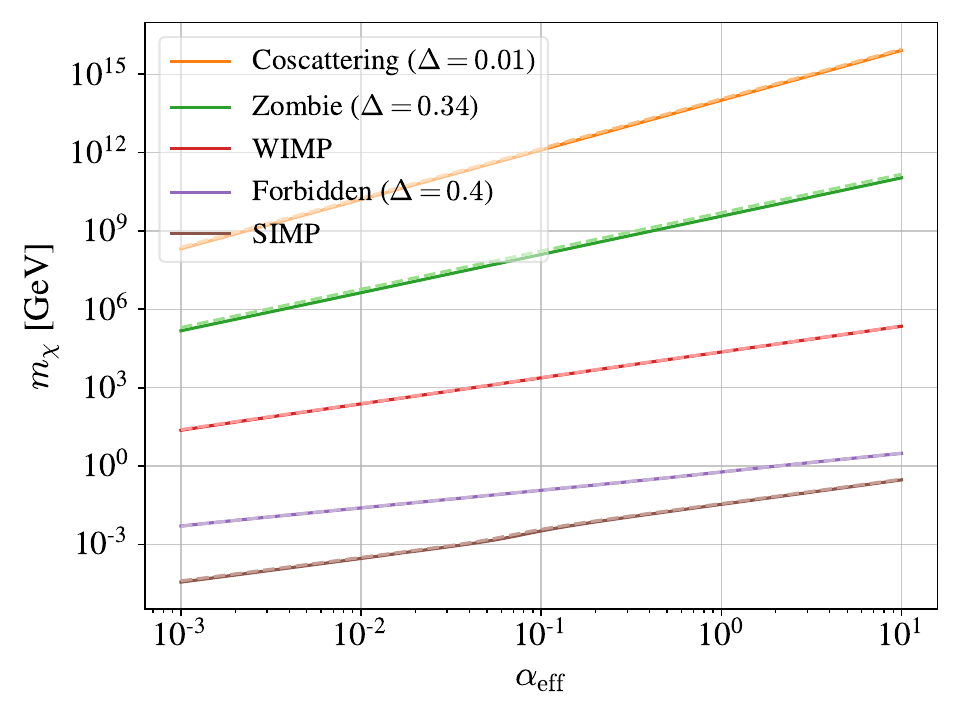}
\vspace{0.7cm} 
     \caption{ {\bf Left:} $m_\chi Y_\chi$ vs. $x$ for DM freezeout. The orange curves shows the solution of the Boltzmann equation for a $m_\chi=1$~GeV WIMP. The red point shows the value $x_d$, given by Eqs~\eqref{xd1} and \eqref{xdslow}. After decoupling, the abundance continues to reduce by a factor $e^{-x_d(\beta-1)}$ until it completely freezes out to a final yield, $m_\chi Y_\chi(\infty)= 0.55 T_{\rm eq}$. In gray we also show the abundance curves for  for a $m_\chi=1$~GeV  DM candidate whose relic abundance is determined by a coscattering or $3\to2$ (SIMP) freezeout process. For coscattering, decoupling happens very early, freezeout is slow and $\beta$ is quite large. On the other hand, SIMP DM freezes out very fast and instanaeous freezeout works as a better approximation of the final relic abundance. 
     {\bf Right:} Mass vs. coupling relationship that matches the observed abundance. The solid line shows the results from the numerical simulation, while the dashed line shows the analytical solution in Eq.~\eqref{relicformula2}. The definitions of $\Delta$ for co-scattering, zombie, and forbidden annihilations are given in Table~\ref{tablesigma}.}
    \label{fig:schematic}
\end{figure*}

Defining the  yield $Y={n}/{s}$, the Boltzmann equation takes the familiar form
\beq
\label{boltzy}
\frac{d Y_\chi}{d x} = - \frac{\langle\sigma v\rangle s^{k-1}}{x H} (n_{i_1}^{\rm eq}\cdots n_{i_{n-k}}^{\rm eq}) \left(Y_{\chi}^{k}- Y_{\chi}^{\ell}\left(Y_{\chi}^{eq}\right)^{k-\ell}\right)\,.
\eeq
Typically, this is the form of the Boltzmann equation that is solved numerically for the asymptotic value of the DM abundance, $Y_{\chi}(\infty)$. 
An estimate for the relic abundance of $\chi$ can be obtained by the instantaneous freezeout approximation. In this approximation, it is assumed that the DM density departs from equilibrium and instantly stops annihilating, where its co-moving density freezes to the value when it departed equilibrium. Departure from equilibrium, which defines the temperature $x_d$, occurs roughly when\footnote{{Note that the definition of $x_d$ varies in the literature. In self annihilation scenarios ($n=k=2$) it is common to define $x_{d}$ by $H(x_d) = c \langle\sigma v\rangle n_{\chi}^{\rm eq}$,
where $c$ is chosen such that the instantaneous freezeout approximation provides the best fit to numerical solution of the Botlzmann equations \cite{Kolb:1990vq,Griest:1990kh}. In this \textit{Letter} we focus on $k=1$ process, where the definition in Eq.~\ref{eq: inst.fo} gives better results (\textit{i.e.} Ref.~\cite{Frumkin:2021zng,Kramer:2020sbb}).
Such choice of definition is motivated analytically  by the $Y_{\chi}$ prefactor in Eq.~\ref{boltzy}, and gives close results both for the $k=1$ and $k>1$ cases, as shown in the right panel of Fig~\ref{fig:schematic}.}}
\beq \label{eq: inst.fo}
    \frac{\Gamma (x_d)}{H(x_d)x_d} = 1.
\eeq
The annihilation rate is given as:
\beq
{\Gamma (x)}   = { (n_{i_1}^{\rm eq}\cdots n_{i_{n-k}}^{\rm eq})  (n^{\rm eq}_\chi)^{k-1} }{}\langle\sigma v\rangle \equiv \alpha_{\rm eff}^{m+n-2}m_\chi \frac{ a e^{-bx} }{x^c} 
 \label{ratedef}
\eeq
for some constants, $a$, $b$, and $c$, which are defined by this equation. The values $b$ in different freezeout processes studied in the literature can be seen in Table~\ref{tablesigma}. For instance, for WIMP like $\chi$-$\chi$ annihilations, using Eqs.~\eqref{xs-param} and \eqref{ratedef}, we see that $b=1$ and  $c={3}/{2}$ corresponds to s-wave annihilation while $c={5}/{2}$ corresponds to p-wave annihilation.

Requiring that the co-moving density at the time of freezeout matches the one observed today by Planck~\cite{Aghanim:2018eyx} yields the constraint 
\beq
Y_{\chi}(\infty) = 0.55\times \frac{T_{\rm eq}}{m_\chi} \, , \label{relicvalue}
\eeq 
where $T_{\rm eq}=0.8$~eV is the temperature at matter-radiation equality. Assuming instantaneous freezeout, namely that the DM density abruptly changes from an equilibrium density to a fixed co-moving density at $x=x_d$, then
\beq
{Y_{\chi}(\infty)=Y^{\rm eq}_{\chi}(x_d)} = \frac{45 g_\chi }{2^{5/2} \pi ^{7/2} g_{\star s}}  x_d^{3/2} e^{-x_d} \label{instfo}
\eeq
for a DM particles freezing out while non-relativistic; $g_{\star}$($g_{\star s}$ ) is the relativistic (entropy) degrees of freedom and $g_{\chi}$ is the number of $\chi$ internal degrees of freedom. 
 Using Eqs.~\eqref{eq: inst.fo} 
-- \eqref{instfo}, one finds that the mass-coupling relationship that gives the observed abundance is
\beq
m_\chi \simeq 
\left( \frac{ \alpha^{n+m-2}_{\rm eff} m_{\rm pl} T_{\rm eq}^b}{x_d^{c+\frac{3}{2}b-1}}\right)^{\frac{1}{1+b}} \label{relicformula1}\,\change{,}
\eeq
where $m_{\rm pl}= 2\times 10^{18}$~GeV is the reduced Planck mass and $x_d$ is the solution to
\beq
x_d \simeq \frac{1}{b+1}\log\left(
 \frac{\alpha_{\rm eff}^{m+n-2}m_{\rm pl}}
 {  x_d^{c-\frac{5}{2}}T_{\rm eq}}
\right) \, . 
\label{xd1}
\eeq
Overall numerical factors are left out for simplicity.  These factors can be recovered here and later with the replacement  
\bea
m_{\rm pl } &\to& a\sqrt{\frac{90}{\pi^2 g_\star(x_d)}}m_{\rm pl }\,,\nn\\ T_{\rm eq } &\to& {\frac{\pi^2 (2\pi)^{3/2}g_{\star s }(x_d)}{30(1+\Omega_{\rm b}/\Omega_{\rm DM})g_\chi}} \frac{g_{\star}(T_{eq})}{g_{\star s}(T_{eq})} T_{\rm eq }\,. \label{replacements}
\eea
where $\Omega_{\rm B}$ and $\Omega_{\rm DM}$ are the observed baryonic and DM relic fractional densities, respectively. The factor is typically an $\mathcal{O}(1-10)$ correction to the mass-coupling relationship.

Given the observed DM energy density, the relic abundance is mostly determined by the mass, coupling, and the degree of  exponential suppression of the annihilation rate, $b$. From Eq.~\eqref{relicformula1} one can obtain as particular examples the mass-coupling relationships for well-studied scenarios such as WIMPs~\cite{Lee:1977ua,Scherrer:1985zt}, SIMPs~\cite{Hochberg:2014dra}, forbidden~\cite{Griest:1990kh,DAgnolo:2015ujb}, co-annihilations~\cite{Griest:1990kh}, co-scattering~\cite{DAgnolo:2017dbv,Kim:2019udq}, zombie~\cite{Berlin:2017ife,Kramer:2020sbb,Bian:2021vmi} and inverse decay processes~\cite{Garny:2017rxs,Frumkin:2021zng}, which are given in Table~\ref{tablesigma}.  

Instantaneous freezeout is a good approximation in many scenarios, because the annihilation rate is dropping exponentially with time (or temperature). However, when the exponential suppression is  weaker ({${b < 1}$}),  instantaneous freezeout may no longer be a good approximation for estimating the relic abundance, as shown in %\sout{. This will often be the case when going beyond the WIMP unitarity bound, which requires $b<1$. Indeed, it was shown in}  
 Ref.~\cite{Kim:2019udq,Kramer:2020sbb}. % \sout{that  freezeout can be very slow.}
  For $b \ll 1$, corresponding to an annihilation with a much lighter particle, the correction can be large. 

Using the parameterization above, $b<1 $  is possible only when $k=1$. Therefore, we consider here corrections to instantaneous  freezeout when there is only one DM particle in the initial state, although the method here can be applied to any scenario. 
The Boltzmann equation~\eqref{boltzy} for $k=1$, using the definitions above, can be written as
\begin{equation}
\frac{d Y_\chi}{d x} = -  \frac{a \alpha_{\rm eff}^{m+n-2} e^{-bx} m_{\chi}}{x^{c+1} H}  \left(Y_{\chi}-Y_{\chi}^{eq}\right)\,.
\label{boltzy_k=1}
\end{equation}
There exists an integral solution to this Boltzmann equation. However, an easy way to get a good estimate for the correction to Eq.~\eqref{relicformula1} is to evolve the equation from the moment $\chi$ departs equilibrium, which we approximate as occurring at $x=x_d$. For $x>x_d$ the inverse reaction can be dropped. The solution is
\beq
Y_\chi (\infty)= Y^{\rm eq}_\chi(x_d) \exp \left( - x_d \int_1^{\infty } x^{-c+1} e^{-b x_d(x-1) } \, dx \right)\,.
\label{y_with_beta}
\eeq
We define a parameter $\beta$ as a measure of the deviation from instantaneous freezeout as 
 \beq
Y_\chi (\infty)
\equiv Y^{\rm eq}_\chi(x_d) e^{-x_d(\beta-1)} \, .  \label{betadef}
\eeq
For $\beta=1$, the instantaneous freezeout approximation is perfect,
while Eq.~\eqref{y_with_beta} provides an additional possible exponential suppression compared to the instantaneous freezeout case given in Eq.~\eqref{instfo}, where  
\beq
\beta = 1+\int_1^{\infty }dx\, x^{-c+1} e^{-b x_d(x-1) } \label{betavalue}.
\eeq

In this form, the mass-coupling relationship in Eq.~\eqref{relicformula1}, can easily be found as before, but using Eq.~\eqref{betadef} instead of Eq.~\eqref{instfo}. One finds
\beq
m_\chi \simeq 
\left( \frac{\alpha^{n+m-2}_{\rm eff} m_{\rm pl} T_{\rm eq}^{b/\beta}}{x_d^{c+\frac{3}{2}\frac{b}{\beta}-1}}\right)^{\frac{1}{1+b/\beta}} \label{relicformula2}\,.
\eeq
where 
\beq
x_d \simeq \frac{1}{b+\beta}\log\left(
 \frac{\alpha_{\rm eff}^{m+n-2}m_{\rm pl}}
 {  x_d^{c-\frac{5}{2}}T_{\rm eq}}
\right).\label{xdslow}
\eeq 
Again, overall numerical factors can be restored with the replacement Eq.~\eqref{replacements}.
Notice that Eq. \eqref{betavalue} and Eq.~\eqref{xdslow} still provide two equations needed to solve for $x_d$ and $\beta$. However, the integral $\beta$ can usually be done semi-analytically. 
For $b\gtrsim 1$, where instantaneous freezeout it expected to be good, one finds that $\beta \simeq 1 +1/(b x_d)$. For $b x_d \ll 1$ (and further more for $c\to 0$), $\beta$ can be quite large. In the extreme $b\to 0$ case, $\beta \simeq 1 +1/(c-2)$. Note that for $b=0$ and $c\le 2$, freezeout never occurs, because the annihilation rate is decreasing slower relative to the Hubble rate; this occurs for decay processes and can also occur for co-scattering with a massless mediator. The values of $\beta$ for various processes are given in Table~\ref{tablesigma}.\footnote{We also include values of $\beta$ for WIMPs and SIMPs and forbidden, still defined by Eq.~\eqref{betadef}. Here $\beta$ is calculated in the same way, but starting with the Boltzmann equation Eq.~\eqref{boltzy}, and dropping the inverse annihilation term for  $x> x_d$.}

In left panel of Fig.~\ref{fig:schematic}  we show a schematic of freezeout, exemplifying the meaning of the different parameters defined in this section. In the right panel of  Fig.~\ref{fig:schematic}  we show the mass vs coupling relationship that matches the observed abundance. The solid line indicates the results from the numerical simulation, while the dashed line indicates the analytical solution in Eq.~\eqref{relicformula2}. The numerical and analytical results are in excellent agreement across many different freezeout processes.

\section{Unitary bound}
\label{sec:unitarity}

% \il{I don't understand the relation between the 2 changes}

Partial-wave  unitarity can be used to derive upper limits on the size of cross-sections. Applying this to DM-DM annihilations, Ref~\cite{Griest:1989wd} showed that partial-wave unitarity gives an upper limit on  the mass of the dark matter from thermal freezeout. The partial-wave  unitarity bound on the cross-section for an arbitrary $n \to m$ process can be found, and then extended to other freezeout processes. However, partial-wave unitarity bound on cross sections is similar in magnitude to the perturbitivity bound on the couplings. Therefore, we simply use perturbitivity as a proxy for partial-wave unitarity, requiring the maximum coupling be $\alpha_{\rm eff}\sim 4\pi$.

From Eq.~\eqref{relicformula1}, the perturbative unitarity bound for WIMP-like DM-DM annihilations is immediately apparent by substituting %\sout{. For DM-DM annihilations into two light bath particles, we have} 
$b=1$, $m=n=2$, and $\alpha_{\rm eff} \sim \mathcal{O}(10)$: 
\beq
m_{\rm WIMP} \simeq \alpha_{\rm eff} \left( m_{\rm pl} T_{\rm eq}\right)^{1/2}  \lesssim 300~\rm TeV. \label{wimpbound}
\eeq 
For perturbative couplings, one sees that the mass of a WIMP DM candidate is never greater than a few hundred TeV. However, going beyond this mass is still possible for a thermal relic, provided that $b<1$---in other words,  that the annihilation rate is less exponentially suppressed  as the DM density is annihilating away.\footnote{There are another two  interesting possibilities to get an exponentially fast rate, but they require species to be out of chemical equilibrium. One is to consider enhancement via stimulated emission if the DM annihilates to a boson with large occupation number. Another possibility is if DM annihilates with a particle that has already frozen out and therefore has a large chemical potential. See Refs.~\cite{Garny:2017rxs,Dror:2016rxc,Kramer:2020sbb,DAgnolo:2017dbv,Maity:2019hre} for freezeout mechanisms that involve particles chemically decoupled from the bath. }   The unitarity bound for an arbitrary interaction  is given by
\beq
m_\chi \lesssim m_{\rm pl} \left(\frac{T_{\rm eq}}{m_{\rm pl}}\right)^{\frac{1}{1+\beta/b}} {x_d^{-\frac{c-1}{1+b/\beta}}}. \label{unitarityb}
\eeq 
Thus for sufficiently small $b$ and $c\le 1$, one can expect that there might be a perturbative thermal candidate for DM all the way up to the Planck scale. 

As we will now show, going above the perturbative WIMP unitarity bound generically implies that there are additional nearly degenerate particles to the DM, and that the DM is metastable. To go beyond WIMP unitarity, first consider the process
\beq
\chi + i_1 + \cdots +i_{n-1} \to f_1 + \cdots +f_m \
\eeq
setting the abundance. %\footnote{We assume that this is the dominant process, which itself can impose hierarchy on the masses of the particles involved.} 
We need only consider the case with one DM in the initial state, otherwise $b\ge1$. The rate for this process scales as 
\beq
\Gamma \propto 
 	\exp \left[-\begin{cases}
{\frac{\sum_i m_i}{m_\chi} x}&  \scriptstyle m_\chi +\dsum_i m_i>  \dsum_f m_f\\
 	{\frac{\sum_f m_f-m_\chi}{m_\chi} x}&  \scriptstyle m_\chi +\dsum_i m_i \le \dsum_f m_f
 \end{cases} \right] \label{unirate}
\eeq
where the second case is for forbidden DM. Requiring $b<1$ imposes
\beq
  \sum_i m_i < m_\chi~{\rm and }~  m_\chi +\sum_i m_i>  \sum_f m_f \label{constr1}
\eeq
or
\beq
  \sum_f m_f < 2 m_\chi~{\rm and }~  m_\chi +\sum_i m_i\le  \sum_f m_f. \label{constr2}
\eeq
 Finally, we require that the DM is stable from the decay induced by moving all initial particles to the final state
 \beq
\sum_i m_i + \sum_f m_f \ge m_\chi . \label{lhsy}
\eeq

\begin{figure}[t!]
    \centering
\includegraphics[width=\columnwidth]{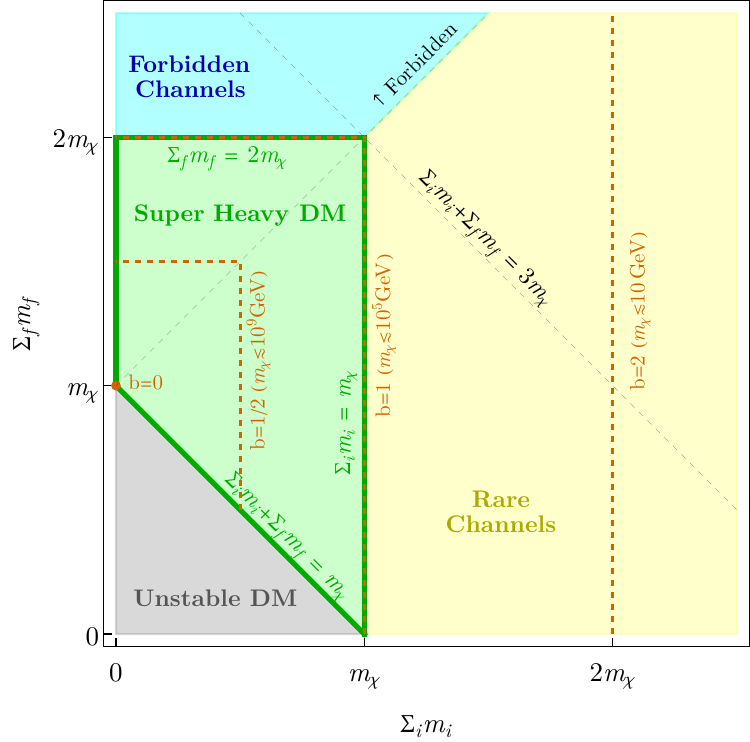}
     \caption{Parameter space for thermal DM freezeout for the general process $\chi + i_1 + \cdots+ i_{n-1} \to f_1  +\cdots +f_m $. Here $\sum_i m_i$ is the sum of the initial particle masses excluding one $\chi$ particle, and $\sum_f m_f$ is the sum of the final particle masses. The green region shows the parameter space given by Eqs.~\eqref{constr1} and \eqref{constr2} where $b<1$ and dark matter can be above the WIMP unitarity bound. The gray region shows where Eqs.~\eqref{lhsy} is not satisfied and the $\chi$ is expected to decay too fast to be the observed DM. The blue and yellow regions show where $b>1$, corresponding to `Forbidden Channels' where the total mass in the final state is heavier than the initial state, and `Rare Channels' where the DM is annihilating off of particles (cumulatively) rarer than itself. Also shown are contours (red-dashed) of constant $b$ for $b=0,0.5,1,2$, and the corresponding upper bound on the DM mass. The upper bound is found is using $\alpha=10$ and $c=3/2$. }
    \label{fig:phases}
\end{figure}

These constraints are shown in Fig.~\ref{fig:phases}. The green area shows the region where DM can be above the WIMP unitarity bound. A simple constraint that can 
be found by combining these conditions gives %\footnote{For example, the l.h.s is given by eq.~\eqref{lhsy}, while for the r.h.s one can take the right part of eq.~\eqref{constr1} and to it the left part of eq.~\eqref{constr1} and finally use the left part of eq.~\eqref{constr1} to get $3m_\chi$.} 
one of the main results of this \textit{Letter}, 
\beq
m_\chi \le \sum_i m_i + \sum_f m_f < 3 m_\chi. \label{mainresult}
\eeq 

Although Eq.~\eqref{mainresult} is not the strongest constraint that can be derived (see Fig.~\ref{fig:phases}), a direct result of Eq.~\eqref{mainresult} %and~\eqref{Finalineq} 
%is that in order for a thermal DM freezeout mechanism to exceed the WIMP unitarity bound, and provide a superheavy stable or metastable DM, there must be a chain of particles with hierarchy of masses, as we explain below. 
is that if the DM mass is beyond the WIMP unitarity bound, then there is another particle close to its mass, which we will call $\chi'$\footnote{It is possible that there is a very large number of particles in the process,  possibly avoiding a degeneracy. However, this would give a large phase space suppression in the cross-section which would ultimately reduce the DM mass.}. 
Since $\chi'$ mass exceeds the WIMP unitarity bound as well ($ m_{\chi'} \sim m_{\chi} $), its abundance will be too large, unless it has some process that efficiently depletes its abundance. If it is unstable, then it may potentially cause the DM itself to be unstable.  If it is stable, it must annihilate away via a non-WIMP like process, and Eq.~\eqref{mainresult} applies to $\chi'$ as well, resulting in a second particle, $\chi''$, with mass close to $
\chi'$ ($ m_{\chi''} \sim m_{\chi'} $). Repeating this argument iteratively for $\chi''$ and beyond, leads to a chain of DM particles and interactions. The chain will end  either when  the mass of last particles goes below the WIMP unitarity bound, or when it reaches a particle that decays in equilibrium with the bath. In the former case the DM may be absolutely stable, but in the the latter case, the DM will be metastable.

As we have shown, a DM chain is a generic consequence of stable or metastable thermal superheavy DM within standard cosmological history. In the next section, we present two examples of such DM chains: a zombie-type chain that gives raise to  stable DM, orders of magnitude heavier than the WIMP unitarity bound, with only small number of particles in the chain. We include a second example of an inverse decay chain, which result in metastable DM with very high masses.

Eq.~\eqref{mainresult} gives a constraint on the masses of the particles involved in the freezeout process, when we require that the DM mass is above the WIMP unitarity bound ($b=1$). A stronger constraint can be obtained if we require that the DM mass be even heavier. For instance, we can impose that the DM is heavier than the mass given by the unitarity limit in Eq.~\eqref{unitarityb} for arbitrary $b$. In Fig.~\ref{fig:phases}, the dashed orange curves show contours of constant $b$, and their corresponding unitarity bound on the dark matter mass.

\section{Freezeout chains}
\label{sec:chain}

Having established a roadmap to going beyond the WIMP unitarity limit for thermal DM, we now turn to new mechanisms that predict super heavy DM. 
We will focus on two chain mechanisms allowing for orders of magnitude larger mass than the WIMP unitarity bound. First we discuss a zombie chain, in which absolutely stable DM is possible well beyond the WIMP unitarity bound with only a small chain. Next we study   an inverse decay chain, where DM is metastable, but whose mass may be much larger.

\begin{figure*}[t!]
  \centering
\includegraphics[width=1\columnwidth]{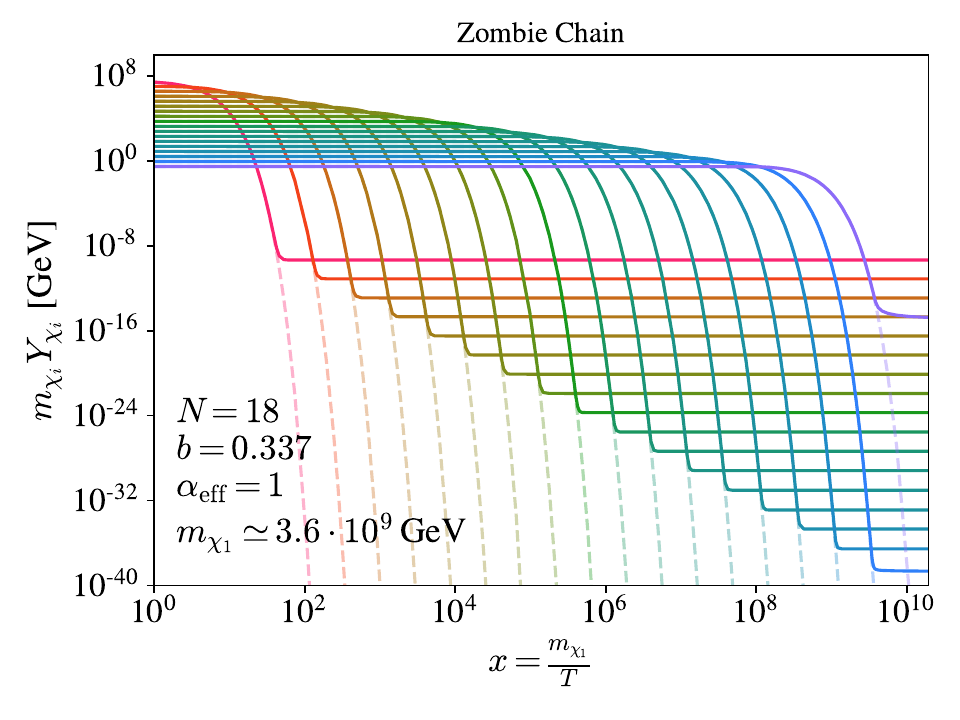}\hfill \includegraphics[width=1.0\columnwidth]{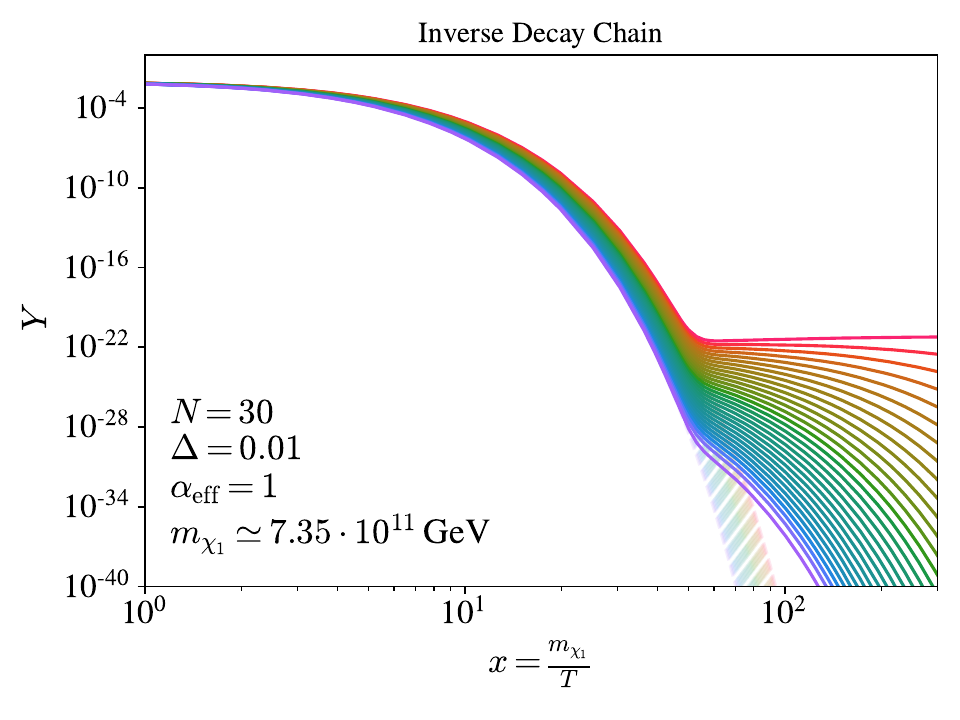}
    \caption{Thermal evolution of the comoving energy density $m_{\chi_i} Y_{\chi_i}$ vs. temperature $x=m_{\chi_i}/T$ for the zombie chain ({\bf left}) and inverse decay chain ({\bf right}). The solid lines shows the numerical abundance for each particle, while the dashed lines represent equilibrium abundance curves.  In each case, we indicate the DM mass, mass-splitting, effective coupling and number of particles in the chain on the plot. }
    \label{fig:indychain}
\end{figure*} 

\subsection{The Zombie Chain}
Next we study a chain based on zombie type-interactions. It was shown in~\cite{Kramer:2020sbb} that a zombie process can allow for a heavy DM candidate without a chain, but with an unstable DM particle. Here we introduce  a  chain of zombie interactions, and show that it supports exponentially larger mass for each additional particle in the chain, up to the bound set by Eq.~\eqref{relicformula2},  while allowing for completely stable DM.

The zombie chain consists of $N$ DM particles, $\chi_i$ ($i=1...N$) with zombie-type nearest neighbor interactions
\begin{equation}
  \chi_i +\chi_{i+1} \leftrightarrow \chi_{i+1} +\chi_{i+1}\,.
\end{equation}
For simplicity, we will assume that each chain interaction has the same strength and that the mass hierarchy is constant going from the heaviest to lightest. Namely, we take $\left< \sigma v \right>_{ i,{i+1} \to {i+1} ,{i+1}}\equiv\alpha^2/m_{\chi_i}^2$ and $m_{i+1} /m_i$  independent of $i$.
The last particle in the chain, $\chi_N$, is assumed to also have direct annihilations into the SM bath particles
\begin{equation}
  \chi_N +\chi_N \leftrightarrow {\rm SM }+{\rm SM }\, . \label{zombieN}
\end{equation}
Since the abundance of $\chi_N$ is determined by standard self-annihilations, and is assumed to be stable, its mass must be less that the WIMP unitarity bound, i.e., $m_{\chi_N} \lesssim 100$~TeV.  

For the zombie interactions, the annihilation rate is 
\beq
\Gamma_{ i,{i+1} \to {i+1} ,{i+1}} = n_{   {i+1}}^{\rm eq} \left< \sigma v \right>_{ i,{i+1} \to {i+1} ,{i+1}} \propto e^{-\frac{m_{{i+1}}}{m_{i}}\frac{m_{{i}}}{T}}
\eeq
and here $b_{\rm zombie}=m_{i+1}/m_{i}$. Following Eq. \eqref{mainresult} and Eq.  \eqref{Finalineq}, we see that in order to go beyond the WIMP unitarity bound, each link in the chain must satisfy
\beq
m_i \leq 3 m_{i+1} = 3 b \, m_i
\eeq
The highest mass is obtained for the smallest allowed value $b=1/3$; below this value, the DM is unstable.  Furthermore, for the interactions above, each $\chi_i$ is absolutely stable, although each will have an exponentially smaller relic abundance than the previous.

Precise solutions to the relic abundance can be obtained by solving the coupled Boltzmann equations. However, the freezeout occurs sequentially: first $\chi_1$ freezes out from $\chi_2$, then $\chi_2$ freezes out from $\chi_3$, and so on. At each stage  each particle freezes out with its neighbors still in chemical equilibrium. For each $\chi_i$, we can simply use Eqs.~\eqref{relicformula1} and \eqref{xd1} to determine the abundance. 

For $b=1/3$, the maximum allowed mass for $\chi_1$ is given by Eq.~\eqref{unitarityb}, which gives
\beq
m_{\chi_1} \lesssim 10^{11}~\rm GeV
\eeq
(for $\alpha_{\rm eff} =10$).
In terms of the last particle in the chain, the ${\chi_1}$ mass is $m_{\chi_1} = 3^{N-1} m_{\chi_N} $.  At least $N=13$ particles are required in the chain to achieve $m_{\chi_1}\simeq 10^{11}~\rm GeV$, if $m_{\chi_N}\simeq 100~\rm TeV$. 

In the left panel of Fig.~\ref{fig:indychain}, we show the numerical solution for $N=18$, $b =0.337$, $\alpha_{\rm eff}=1$, and ${m_{\chi_1}=3.6\times 10^{9}}$~GeV, which exceeds the WIMP unitarity bound Eq.~\eqref{wimpbound}. The thermal evolution of the comoving density is shown for each $\chi_i$. The abundances of $\chi_1$ through $\chi_{N-1}$ freezeout via zombie processes and have decreasing relic abundance $m_{\chi_i} Y_{\chi_i}/ m_{\chi_{i+1}} Y_{\chi_{i+1}}\simeq 70$.   The abundance of $\chi_{N}$ freezes out via self annihilations Eq.~\eqref{zombieN}, and therefore has a relatively large relic abundance, but much smaller than the ${\chi_1}$ abundance.

\subsection{The Inverse Decay Chain}

Consider freezeout of $\chi$ via inverse decay (INDY) process~\cite{Garny:2017rxs,Frumkin:2021zng} 
\beq
\chi +\gamma^\prime \to \psi,
\eeq 
where $\gamma^\prime $ is a light particle in equilibrium with the bath. It is easy to see from detailed balance  that the rate for depleting  $\chi $ is 
\beq
\Gamma_{\chi + \gamma^\prime \to \psi} = \frac{n_{\psi}^{\rm eq}}{n_{\chi}^{\rm eq}}  \Gamma_{\psi\to \chi+ \gamma^\prime} \propto  e^{-\frac{m_\psi-m_\chi}{m_{\chi}}x}
\eeq
where $\Gamma_{\psi  \to \chi + \gamma^\prime}$ is the partial width of $\psi$. For inverse decays, in terms of the definition of the rate in Eq.~\eqref{ratedef}, 
\beq
b=\frac{m_\psi-m_\chi}{m_\chi}\equiv \Delta, \qquad c=0,
\eeq  
and 
\beq
\alpha_{\rm eff}= \frac{\Gamma_{\psi\to \chi+ \gamma^\prime} }{m_\chi} (1+\Delta)^{5/2}\,.
\eeq
The suppression of the rate in temperature is only due to the mass-splitting between $\psi$ and $\chi$. Therefore, the rate only  decreases when the mass splitting is of order the temperature,  i.e. $x\Delta>  1$. For sufficiently small $\Delta$, decoupling and freezeout can occur  arbitrarily late. For these reasons it is expected that arbitrarily high DM masses are possible. 

However, for the reasons discussed in the previous section, if $m_\chi$ is well above the WIMP unitarity bound, then there needs to be a way for $ \psi$ to deplete in order to not be over abundant. If it is unstable while decaying in equilibrium, then certainly $\chi$ will be too short lived to be DM. Thus, we need to consider $ \psi$ depleting by inverse decays as well. To this end, we consider a chain of inverse decays in order to achieve arbitrarily high DM mass. The chain must be long enough so that the phase space in the $\chi$ decay is small enough to stabilize the DM on cosmological scales.

Consider the chain of inverse decay processes 
\beq
\chi_i +\gamma^\prime \to \chi_{i+1},
\eeq 
for $i=1\ldots N$, and the decay for $\chi_N$ directly into the Standard Model bath 
\beq
\chi_N \to {\rm sm}+ {\rm sm}.
\eeq 
The decay rate $\chi_N$ is taken to be fast, so that it is an equilibrium process, i.e., $\Gamma_{\chi_N \to {\rm sm}+ {\rm sm}} >H(m_{\chi_N})$. 
For simplicity we consider the case where the decay width for each process in the chain is constant ${\Gamma\equiv\Gamma_{\chi_{i+1} \to \chi_i + \gamma'} }$ and a constant mass splitting ${m_{\chi_{i+1}} = m_{\chi_i}(1+\Delta)}$.

The relic abundance can be found by solving the $N$ coupled Boltzmann equations: 
\bea
xH \frac{\partial Y_{1}}{\partial x} &=& {\Gamma}{} \left(Y_{2} - Y_{1} \frac{n_{{2}}^{\rm eq}}{n_{1}^{\rm eq}} \right) \\
&\vdots &\nn\\
xH\frac{\partial Y_{i}}{\partial x} &=& {\Gamma}{}  \left( Y_{{i+1}} - Y_{i} \frac{n_{{i+1}}^{\rm eq}}{n_{i}^{\rm eq}} \right) +{\Gamma}{}  \left( Y_{{i-1}} \frac{n_{{i}}^{\rm eq}}{n_{{i-1}}^{\rm eq}} - Y_{{i}} \right) \nn\\
&\vdots &\nn\\
xH\frac{\partial Y_{N}}{\partial x} &=& {\Gamma}{}  \left( Y_{{N-1}} \frac{n_{{N}}^{\rm eq}}{n_{{N-1}}^{\rm eq}} - Y_{{N}} \right)- {\Gamma_{N \to {\rm sm}}  } \left( Y_N-Y_N^{\rm eq}\right)\,,\nn
\eea
where $x= m_{\chi_1}/T$.

We show a numerical solution in the right panel of Fig.~\ref{fig:indychain}~ for $N=30$, $\Delta =0.01$, $\alpha_{\rm eff}=1$, and $m_{\chi_1}= 7.35\times 10^{11}$~GeV, which exceeds the WIMP unitarity bound, Eq.~\eqref{wimpbound}. One can see that the analytic mass-coupling relationship derived in Eq.~\eqref{relicformula2} does not well-describe the results here. Similar to what was observed for the coscattering chain in Ref.~\cite{Kim:2019udq}, the chain can weaken the strength of the DM depletion. For scattering, it was shown that the effective rate for annihilation is suppressed by $N^2$. In other words, one should replace  $\alpha_{\rm eff}\to \alpha_{\rm eff}/N^2$  in order to obtain an estimate for correct coupling needed to match the observed abundance. Exact analytic solutions to the coupled set of equations are difficult to obtain, and we leave this to future work~\cite{inversechain}. 

Finally, we comment on the stability of the DM for the example shown. The DM candidate, $\chi_1$, can decay to $N$ SM particles via $N-1$ off-shell $\chi_i$ particles. We estimate the phase-space and choose $N=30$ as a safe value that guarantees that the phase space in the $\chi$ decay is small enough to stabilize the DM on cosmological scales.

\section{Outlook}
\label{sec:outlook}
In this \textit{Letter} we have derived general formula for the thermal relic abundance of dark matter for an arbitrary freezeout process, and have outlined the roadmap to obtain a thermal relic with mass above the standard WIMP perturbative unitarity bound of a few hundred TeV.

Thermal freezeout provides a well-motivated framework for studying the relic abundance of DM. The relic abundance can be determined in the instantaneous freezeout approximation for most cases with Eq.~\eqref{relicformula1}, or with Eq.~\eqref{relicformula2} which corrects for when freezeout is slow. Depending on the process controlling freezeout, there is a different bound on the dark matter mass from perturbative unitarity, which can be substantially higher than that of the WIMP standard lore. To go beyond the standard WIMP unitarity bound, we find  that there are  particles degenerate with the DM,  possibly constituting a long chain of DM interactions. 
Additionally, if the DM is a heavy thermal relic within a standard cosmological history, then generically it is expected be meta-stable. This would lead to a striking signal of ultra-high-energy cosmic rays (UHECR) from decaying DM, that can be searched for in dedicated UHECR detectors such as Ice Cube~\cite{IceCube:2022clp} and  Pierre Auger Observatory~\cite{PierreAuger:2022uwd} (in addition to gamma-ray satellites such as FERMI-LAT~\cite{Fermi-LAT:2014ryh}).

Much work remains to be done to better understand heavy dark matter in context of a chain and to populate the model parameter space that achieves such heavy thermal relics. We leave this to future work~\cite{inversechain}.

\begin{acknowledgments}
{\bf Acknowledgments.}
We thank Hyun Min Lee and Josh Ruderman for useful discussions. We are especially grateful to Yonit Hochberg and Nadav Outmezguine for useful discussions and comments on the manuscript. 
The works of RF, EK, IL, and TS are supported by the US-Israeli Binational Science Foundation (grants 2016153 and 2020220), the Israel Science Foundation (grant No. 1111/17), and by the I-CORE Program of the Planning and Budgeting Committee (grant No. 1937/12).
\end{acknowledgments}

\bibliographystyle{apsrev4-1}
\bibliography{biblio}{}

\end{document}